\def\beginEq{\begin{equation}}
\def\endEq{\end{equation}}
\def\beginEqarray{\begin{eqnarray}}
\def\endEqarray{\end{eqnarray}}

\def\nl{\nonumber \\}

\def\NRQED{{\rm NRQED}}

\def\Lag{{\cal L}}    

\def\Dv{{\bf D}}
\def\sigmav{\mbox{\boldmath$\sigma$}}
\def\Ev{{\bf E}}
\def\Bv{{\bf B}}
\def\psid{{\psi^\dagger}}
\def\chid{{\chi^\dagger}}

\def\quarter{\mbox{$\frac{1}{4}$}}
\def\DIV{{\rm div}}

\def\boxit#1{\vbox{\hrule\hbox{\vrule\kern3pt
\vbox{\kern3pt#1\kern3pt}\kern3pt\vrule}\hrule}}

\def\expect#1{{ \langle{#1}\rangle }}

\def\twowave{ \int {d^3p d^3q \over (2\pi)^6}~ {(8 \pi^{1/2}
\gamma^{5/2})^2
\over ( {\vec p}^2 +\gamma^2)^2 ( {\vec q}^2 + \gamma^2 )^2 }}
\def\transv{ {-i \over ( \vec p - \vec q)^2 } ( \delta_{ij}-{ (\vec p -
 \vec q)_i (\vec p - \vec q)_j  \over (\vec p - \vec q )^2 } ) }

\def\frac#1#2{{#1}\,/{#2}}
\def\psid{{\psi^\dagger}}
\def\chid{{\chi^\dagger}}
\def\Bv{{\bf B}}
\def\Ev{{\bf E}}

\def\quarter{{\frac{1}{4}}}

\def\Dv{{\bf D}}
\def\DIV{{\bf \nabla}}
\def\g5{{\gamma_5}}

\def\A0{{A_{1/2}}}
\def\A2{{A_{3/2}}}

\def\vec#1{{\bf #1}}

\def\be{\begin{equation}}
\def\ee{\end{equation}}
\def\ba{\begin{eqnarray}}
\def\ea{\end{eqnarray}}

\def\centeronto#1#2{{\setbox0=\hbox{#1}\setbox1=\hbox{#2}\ifdim
\wd1>\wd0\kern.5\wd1\kern-.5\wd0\fi
\copy0\kern-.5\wd0\kern-.5\wd1\copy1\ifdim\wd0>\wd1
\kern.5\wd0\kern-.5\wd1\fi}}

\newdimen\pmboffset
\def\oldpmb#1{\setbox0=\hbox{#1}%
 \copy0\kern-\wd0
 \kern\pmboffset\raise 1.732\pmboffset\copy0\kern-\wd0
 \kern\pmboffset\box0}

\documentstyle[12pt]{article}

\title{Higher order corrections to bound state energy levels
in QED: an effective field theory approach}
\author{Patrick Labelle 
  \\ \small Department of Physics, McGill University 
\\ \small 3600 University St., Montr\'eal, Qc H3A-2T8  Canada
}

\begin{document}
\maketitle
\begin{abstract}
In \cite{koniuk}, a new method was developed to calculate energy levels
in QED nonrelativistic bound states, up to order $m \alpha^5$ (or
$\alpha^3$ with respect to the Bohr levels). Whether or not this
method could be used beyond this order was left as an open question. 
In this paper, we answer this question with the help of a
nonrelativistic effective field theory, NRQED. This theory permits
to deal separately with bound state physics (taking place at
momenta of order $ m \alpha$ or below) from relativistic
physics (where the momentum is of order the electron mass) in
a rigorous and systematic manner. We find that, aside from 
infrared terms which had to  be discarded without justification
in \cite{koniuk}, the ${\cal O} (m \alpha^4)$ and ${\cal O} ( m
\alpha^5)$ calculations give the same results in both methods. 
It is shown, however, that this is due to the special physical
origin of these contributions
and that the method of \cite{koniuk} would
not give the right answer at order $m \alpha^6$ or higher.
The separation of scales provided by NRQED is essential to the
derivation of this result.
\end{abstract}

\section{Introduction}
In \cite{koniuk}, fine and hyperfine splittings calculations of nonrelativistic
QED bound states are carried out up to order $m \alpha^5$, or
$\alpha^3$ with respect to the lowest order (Bohr) energy levels. The shifts in
energy are obtained, essentially, by sandwiching Feynman scattering 
amplitudes between Schr\"odinger wavefunctions. These amplitudes are
evaluated with the external particles on mass-shell and their
four-momenta taken in the nonrelativistic limit. The authors left open the
question of whether this approach could be pushed beyond this order.

In this paper, we answer this question using a nonrelativistic effective
field theory, nonrelativistic quantum electrodynamics (NRQED)\cite{NRQED}.
 The use
of such an effective theory is possible because the typical QED
bound state
momentum is of order $\mu \alpha$ (where  $\mu$ is the reduced mass),
which is much smaller than the masses of the constituents. 
 In an
effective field theory,  an infinite number of
nonrenormalizable interactions are present,
but they are suppressed by factors of $ 1 / \Lambda $ where $\Lambda$
is an appropriate cutoff characterizing a scale of energy at which
the theory would break down. Because of the nonrelativistic nature
of the bound states we are considering, it is natural to take this
cutoff as being the reduced mass of the system. It is
important to emphasize that the physics taking place at energies
beyond the cutoff $\Lambda$  {\it does} affect the low energy physics
and thus must not be thrown away. However, because this physics takes
place at such a high energy (in comparison to the  domain of  energy
in which the effective theory is used), it can be incorporated back
in the low energy theory by renormalizing its coefficients.
Let us emphasize that this procedure is performed in a rigorous and
systematic way and that the low energy theory can thus be made to
reproduce QED to an arbitrary precision (as long as the system
under study is characterized by momenta $p < \Lambda$).

We find that the approach of \cite{koniuk} would be inconsistent at the
${\cal O} ( m \alpha^6)$ and would not lead to  the correct shift in energy.
Moreover, using NRQED, we will be able to understand why  this approach
gives the correct order $ ( m \alpha^4)$ and $ m \alpha^5$ 
contributions\footnote{Except for the infrared divergent terms which,
in the formalism of \cite{koniuk}, must be set aside without any
further justification. In the context of NRQED, however, these
terms are systematically removed.}.

This paper
is organized as follows. In Section 1, we explain the difficulties
associated with QED nonrelativistic bound states. In the following two
sections, we show how to build a nonrelativistic effective field theory
and why it is particularly useful for the study of bound states. We then 
evaluate the hyperfine splitting
 in Positronium up to order $m \alpha^5$ and show
what distinguishes the  ${\cal O} (\alpha^4) $
and  ${\cal O} (\alpha^5) $
 terms from higher order corrections. This will clearly show
why the approach of \cite{koniuk} is valid at that order (except for the
infrared cutoff dependence), and also why one can generalize
 the order $ m \alpha^5$ result to
  arbitrary values of n, the principal quantum number.
We conclude by explaining
how to proceed to obtain the $m \alpha^6$ and higher corrections,
and why the approach of Ref.\cite{koniuk} would fail at this stage.

\section{Nonrelativistic bound states}
 In a nonrelativistic bound state,  the traditional expansion
in the number of loops familiar from quantum electrodynamics (QED)
fails. The physical reason is the presence of energy scales absent from
scattering theory. Indeed, the finite size of the atom provides
an energy scale of order the Bohr radius $\approx 1 / m \alpha$ which,
by the uncertainty principle, is equal to the inverse of the typical
three-momentum of the constituent particle ($\approx m \alpha$). Because
of this new energy scale, there is a region of the momentum
integration in which the addition of  loops will
{\it not} result in additional factors of $\alpha$. Also, because
this energy scale is much smaller than the particles masses, the
system is predominantly nonrelativistic, a simplification not
taken advantage of in traditional approaches. Moreover, a third
energy scale, again vastly different from the previous two,  is set
by the particles kinetic energies and further complicates bound state
calculations. It is 
 at the origin of the so-called
retardation corrections which lead, for example, to the Lamb shift.

To illustrate more specifically the problems arising
from the presence of these energy scales,
 consider a traditional
Bethe-Salpeter\cite{BS}  calculation of the energy
 levels of positronium. The potential
in the Bethe-Salpeter equation is given by a sum of two-particle irreducible
Feynman amplitudes. One generally solves the problem for some approximate
potential and then incorporates corrections using time-independent
perturbation theory. Unfortunately, perturbation theory for a bound state
is far more complicated than perturbation theory for, say, the electron
$g$-factor. In the latter case a diagram with three photons contributes
only to order $\alpha^3$. In positronium a kernel involving the exchange
of three photons   contributes not only  to order
$\alpha^3$, but also  to all higher orders
as well\footnote{We assume here that the Coulomb component of each
photon exchange has been projected out. See Section 4 for more
details on the special status of the Coulomb interaction in 
nonrelativistic bound states.}:
\smallskip
\begin{equation}
 \expect{V_1}~=~\alpha^3 m \bigl( a_0 + a_1 \alpha + a_2 \alpha^2+ ...
\bigr) .
\end{equation}
Therefore,  in the bound state calculation there is no simple correlation
between the importance of an amplitude an the number of photons in it.
Such behavior is at the root of the complexities in high-precision
analyses of positronium or other QED bound states, and it is a direct
consequence of the multiple scales in the problem. Any expectation
value like Eq.(1) will be some complicated function of the ratios
of the three scales in the atom:
\smallskip
\begin{equation}
 \expect{V_1}~=~\alpha^3 m~ F(\expect{p}/m,\expect{K}/m).
\end{equation}
\smallskip
Since $\expect{p}/m \sim \alpha$ and $\expect{K}/m \sim \alpha^2$, a Taylor
expansion of $F$ in powers of these ratios generates an infinite series
of contributions just as in Eq.(1).  A similar series does not
occur in the $g$-factor calculation because there is but one scale
in that problem, the mass of the electron.

 Let us again emphasize that traditional 
methods for analyzing these bound states fail to take
advantage of the nonrelativistic character of these systems; and
atoms like positronium are very nonrelativistic: the probability for
finding relative momenta of ${\cal O}(m)$ or larger is roughly $
\alpha^5 \simeq 10^{-11}$!

Let us also point out that the difficulties associated with the
many different energy scales in bound states are not limited to QED. Consider
the study of nonrelativistic QCD bound states,  the $\Upsilon$
for example. Even if the means of extracting the properties of such
systems, namely simulating the system on a lattice, are very different than the
use of the Bethe-Salpeter equation, the nonrelativistic nature of
the bound state still greatly complicates  calculations. Indeed, the
space-time grid used in such a simulation must accommodate wavelenghts
covering all of the scales in the meson, ranging from $(mv^2)^{-1}$ down
to $m^{-1}$. Given that $v^2 \sim 0.1$ in $\Upsilon$, one might easily
need a lattice as large as 100 sites on side to do a good job. Such a lattice
would be three times larger than the largest wavelenght, with a grid spacing
three times smaller than the smallest wavelenght, thereby limiting the errors
caused by the grid. This is a fantastically large lattice by contemporary
standards and is quite impractical.
 The strong force equivalent of NRQED,$\,$  NRQCD\cite{NRQCD}, greatly reduces
the size of the lattices required to study heavy quarks systems
 since it essentially eliminates the mass of the
meson as a relevant energy scale in the calculation (we will soon see
how this occurs in NRQED).

\section{NRQED}

To take advantage of the simplifications associated with nonrelativistic
systems, one can approximate QED in the limit $p/m \ll 1$ (which is
equivalent to the limit $v/c \ll 1$). One way
to do this consists of simply expanding the QED Lagrangian  in powers
of $p/m$ to obtain NRQED. The usefulness of this expansion is enhanced
by performing a Foldy-Wouthuysen transformation, which decouples
the electron and positron degrees of freedom. It is also, however,
possible to define
NQRED without prior knowledge of QED.
Indeed, one can build up the Lagrangian
of NRQED by imposing restrictions coming from the symmetry obeyed by the
theory, such as gauge invariance,  chiral symmetry in the limit $m_e \rightarrow
0$, and Lorentz invariance for the photon kinetic term. Lorentz invariance
for the rest of the Lagrangian and renormalizability are {\it not} necessary.

It is of course possible to write down an infinite number of terms respecting
these symmetries,
but this is not a problem since operators of dimension greater than $4$ will
be multiplied by coefficients containing inverse powers of $m$, the only
available energy scale in the problem (remember that $m$ plays
the role of the ultraviolet cutoff $\Lambda$ in NRQED).
 Since we are considering the
limit  $p/m \ll 1$, these operators  will yield
contributions suppressed by that many powers of $p/m$ and therefore,
for a given accuracy, only a few terms need to be kept in an actual
calculation. The first few terms one obtains are then
\beginEqarray
\Lag_{\NRQED} & = &  - \quarter(F^{\mu\nu})^2 +
  \psid \biggl\{ i\partial_t - e A_0 + \frac{\Dv^2}{2m} + \frac{\Dv^4}{8m^3}
   \nl
&& + c_1~\frac{e}{2m}\,\sigmav\cdot\Bv + c_2\frac{e}{8m^2}\,\DIV \cdot \Ev \nl
&& + c_3~\frac{ie}{8m^2} \,\sigmav\cdot(\Dv\times\Ev - \Ev\times\Dv)
   + ...\biggr\} \psi \nl
&& + d_1 \frac{e^2}{m^2}\,(\psid\psi)^2
 + d_2 \frac{e^2}{m^2}\,(\psid\sigmav\psi)^2
 +... \nl
 && + \mbox{~positron and positron-electron terms}.
 \label{nrlag}
 \endEqarray
where the c's and d's are coefficients of order one.
Here, $\psi$ and $\chi$ are the (two-component) electron and positron
spinors and $\Dv$ is the gauge-invariant derivative. An example of a
 positron-electron term  (and the one we will focus on  later)
 is $ d_3 e^2/m^2 \,
(\psid\sigmav \chi) \cdot (\chid\sigmav \psi) $, which represents, to
lowest order, the process $e^+ e^- \rightarrow e^+ e^- $ in the $s$
channel. This term, and the other two having coefficients starting with
$d$'s are examples of ``contact" interactions.

 Notice  that
we have  recovered the interactions familiar from nonrelativistic quantum
mechanics such as the first order relativistic correction to the
kinetic energy ($ \Dv^4/8m^3$)
 and the Darwin term ($ c_2~e/8m^2\,\DIV \cdot \Ev$).

Of course, one cannot perform a nonrelativistic expansion of the
photon kinetic term. It is however still possible to simplify the
nonrelativistic calculation
 by choosing the most useful
gauge in that context, the Coulomb gauge (the reason for its
usefulness will become clear later on). In that gauge, the
Coulomb propagator is given by $i / \vec k^2$ and the transverse
propagator is given by
\begin{equation}
{i \over k^2 + i \epsilon} \bigl( \delta_{ij} - { k_i k_j \over \vec k^2}
\bigr) .
\end{equation}
The non recoil limit is obtained by setting $k^2 = - \vec k^2$ in the
transverse photon propagator, which amounts to neglecting the energy
transferred by the photon.

We now have to fix the coefficients appearing in $\Lag_{\NRQED}$. 
To do so, we may simply compare  scattering amplitudes in both
QED and NRQED, at a given kinematic point (which we choose to be at threshold,
{\it i.e.} with the external particles at rest). This renormalization
of the low energy coefficients (often referred to as the ``matching
procedure" in the literature on effective field  theories) must be carried
out to as many loops as required for the  accuracy desired
in the final calculation;
 at that stage, one can associate a factor
of $e$ for each vertex to evaluate the contribution to the
bound state calculation (see section [4.1] for more details).
 Only a finite number of NRQED interactions,
for that given number of loops, must be kept.
 That number is determined using  NRQED counting rules, which permit
to determine the order at which a given NRQED diagram will
contribute, {\it after} renormalization (or matching)
has been carried out. We
will  give these rules in  section [4.1].

\section{Bound states diagrams in NRQED}

Now that NRQED is completely defined, we can apply it to bound state
calculations, where its usefulness is  most apparent. To be specific,
consider the energy levels of positronium. In a bound state, the expansion
in Feynman diagrams of NRQED reduces to  conventional Rayleigh-Schr\"odinger
perturbation theory. The ``external" wavefunctions
of course don't correspond to free wavepackets but to bound states.
In the traditional Bethe-Salpeter formalism, the form of these wavefunctions
depend on which interactions one desires to include exactly, and which
interactions
one wants to treat perturbatively.
 Depending on this choice, the external wavefunctions
may be as simple as the Schr\"odinger wavefunctions or as complicated as the
Dirac wavefunctions.

We are now in measure to appreciate the usefulness of the Coulomb gauge.
 This gauge  permits us to keep
 the simplest interaction for the
zeroth order kernel, namely the Coulomb interaction, in which case
the Bethe-Salpeter equation reduces to the Schr\"odinger equation
and the wavefunctions are the ones found in any textbook on quantum mechanics.
In this calculation we will work only with the $1S$ wavefunction in positronium
(reduced mass $\mu=m/2$) which is given, in momentum space, by
\smallskip
\begin{equation}
 \Psi(\vec p)~=~ { 8 \pi^{1/2} \gamma^{5/2} \over
(\vec p^2 + \gamma^2)^2},
\end{equation}
where $\gamma$ is the typical bound state momentum, equal to $\mu v
 = m \alpha /2$ in positronium. The ground state energy is $- \gamma^2/2 \mu =
- m \alpha^2/4 $. Physically, an infinite number of Coulomb interactions is
incorporated in the wavefunction.

 The reason why at least an infinite number  of
Coulomb kernels must be taken into account in the wavefunction can be
easily understood by considering  fig.[1a], where a generic bound state
diagram is shown. $K_1$ and $K_2$ represent some unspecified kernels.
We can schematically represent this diagram by
\begin{equation}
\int {d^3k \over (2 \pi)^3} F(\vec k; \vec k)
\label{integ}
\end{equation}
where $ F(\vec p_1; \vec p_2)$ is a  function including the
bound state wavefunctions, the interactions present in the kernels
$K_1$ and $K_2$, and the propagator of the electron-positron
pair between these two kernels.
  The arguments $\vec p_1$ and $\vec p_2$
represent the three-momenta flowing out of the first kernel and into the
second one. In figure [1a], $\vec p_1= \vec p_2 = \vec k$.
As explained above, working with the effective field theory
means that the electron mass no longer appears as a dynamical
energy scale in the NRQED diagrams; this means that  $m$ can only
appear as an overall factor of the form $1/ m^n$ in figure
 [1a]\footnote{In the language of effective field theories (EFT's),
NRQED is a decoupling EFT with cutoff scale $\Lambda =m$. The 
heavy particle being decoupled is the electron's antiparticle
so that Lorentz invariance is broken (as 
expected in a nonrelativistic theory). Of course, there is a
positron in NRQED, but it is here an independent particle;
it never appears as the negative energy pole of the electron propagator
(in the language of time-ordered perturbation theory, there are no
Z graphs). Consequently, one can apply the decoupling theorem
 \cite{decoupling} to
NRQED, with the consequence that the heavy scale $m$ will
only appear as an overall factor $1/m^n$ in the interactions.
One also expects $m$ to enter in the running of the low energy
coefficients through logarithms, but this dependence on $m$ will
only come from the QED scattering diagrams, through the
matching (see section [5.3].}.
 The function $ F$ is therefore {\it
independent} of the electron mass; the only energy scale it depends
on is the typical bound state momentum $\gamma$, and the ultraviolet
cutoff (when it is divergent)\footnote{We will deal with the infrared
divergences separately.}.

Let us now  consider adding a Coulomb interaction between the two
kernels, as in fig.[1b].
 In conventional
scattering theory, one would conclude that this diagram is of higher
order than  diagram [1a]  because of
the factor $\alpha$ coming from the vertices. However, this argument does
not hold in a bound state, because the typical momentum flowing
through the Coulomb interaction is of order $\gamma$. Indeed, consider
how the integral Eq.[\ref{integ}] is modified when one adds the Coulomb 
interaction.  It must now be written 
\smallskip
\begin{eqnarray}
 && \int {d^3k \,
   d^3l \over (2 \pi)^6}  ~ {i  e^2 \over \vert \vec k - \vec l \vert^2}~
{ i  \over \ (- \gamma^2 - \vec k^2)/m }  F(\vec k; \vec l)
\nonumber \\ && ~=~
4 \pi m \alpha ~   \int { d^3k  \, d^3 l \over ( 2 \pi )^6} { 1
\over (\gamma^2 + \vec k^2) ( \vert \vec k - \vec l \vert^2)} F( \vec k
; \vec l) 
\label{coulomb}
\end{eqnarray}
where $i/ \vert \vec k - \vec l \vert^2$ is the Coulomb propagator
and $ i m/ (- \gamma^2 - \vec k^2)$ is the nonrelativistic propagator for
the $e^-e^+$ pair.

We see that the mass factors out (more about this later!), leaving $\gamma$
as the only energy scale in the integral. By dimensional analysis, the
final result will be of order
 $\simeq m \alpha$ (the overall factor) divided by $ / \gamma $
(from the integral) $\simeq 1$. This shows that in a bound
state, adding a Coulomb interaction  does not, in fact,  lead to a diagram
of higher order. This  is why one must sum up all Coulomb
interactions
in the wavefunctions right from the start.
 This implies the relation expressed in fig.[2]
which is nothing else than a diagrammatic representation of the Schr\"odinger
equation.

\subsection{Counting rules}

The argument we used to show that adding a Coulomb interaction does not
increase the order of a bound state diagram is  an example of a NRQED
counting rule. These rules are extremely important since
 they permit one to estimate the order of
contribution of a diagram without  calculating it.
Let us first consider non-recoil diagrams\footnote{In the case of
a bound state having  equal mass
constituents, as  positronium, there is no distinction possible between
 ``recoil effects" and the so-called ``retardation effects". This is not, for
example, the
case in muonium or hydrogen where (sometimes
confusing) distinctions are made between
non-recoil and non-retardation approximations. What we mean here by the
 non-recoil approximation is the limit in which
 the transverse photon does not carry
any energy.}.
In that approximation, the counting rules are based on the observation that
the mass $m$ of the electron  always factors out of the integrals so that it
 no longer represents a relevant energy scale.
This is clear since $m$ factors out of the nonrelativistic fermion propagators,
as we saw in Eq.[\ref{coulomb}], and it
 enters the rules of the NRQED vertices only
as an overall factor.
We therefore see that  the use
of NRQED eliminates the relativistic energy scale of the system, which, in
the light of the discussion in the introduction,
 greatly simplifies the analysis.

Starting from this observation, we can recast the discussion of the Coulomb
interaction in the following manner: adding a Coulomb interaction between two
kernels as in fig.[1] leads to an additional factor of $\alpha$ coming from
the vertices and a factor of $m$ coming from the additional fermion
propagator. The corresponding overall factor of $m \alpha$ must be canceled
by a factor having the dimensions of energy in order to keep the dimensions
straight, but since the only energy scale left in the problem is $\gamma$,
we finally obtain a correction of order $m \alpha / \gamma \simeq 1 $. It
now becomes obvious that the Coulomb interaction is the only kernel
having the property of not increasing the order of a bound state diagram
since all the other interactions contain additional factors of $1/m$.
 Each of these $1/m$ factors will have to be canceled by a corresponding
factor of $\gamma$, leading therefore to a result of higher order in
$ \alpha$.

With these rules, it becomes extremely simple to evaluate the order of the
contribution of a given bound state diagram.
 All we need to know  is that the external
wavefunctions contribute  a factor $m^3 \alpha^3$. To obtain
the contribution of a given kernel, one has simply to count the number
of
explicit factors of $\alpha$ (coming from the vertices) and inverse
factors
of $m$ (coming from the vertices and the $e^-e^+$ propagators)
this diagram contains.
After canceling the $1/m$'s  with factors of $\gamma$, the 
number of $\alpha$'s left over give the order of the contribution
of the diagram.
 Consider, for example,  inserting
a transverse photon (in the non-recoil limit)  between two wavefunctions
as  in fig.[3]. The transverse photon vertex has one power of
 $e$ and one power of $1/m$.
  Taking the two vertices into account, we can conclude that this
diagram will contribute to order $m^3 \alpha^3 \times \alpha/m^2
  \simeq
m \alpha^4$ (it is not necessary to multiply by any factor of $\gamma$
here). The same is true of the annihilation diagram and of the
spin-spin interaction as they also contain a factor of
$1 /m^2$ relative to the Coulomb interaction. Another important observation
is that these diagrams will contribute to only one order\footnote{We
 are still limiting ourselves to the non-recoil limit.} in $\alpha$,
  in contradistinction with Feynman diagrams in conventional Bethe-Salpeter
analysis. It is important to understand that these rules give the order
of contribution of a diagram {\it after} the renormalization
of the effective theory has been performed.  Indeed, many NRQED
bound state diagrams are actually badly divergent in the ultraviolet.
The rules give us the order of the finite contributions left over
after the  NRQED bare coefficients have been renormalized (or,
in other words, after matching with QED has been performed).

The situation is only slightly more complicated when one takes into account
recoil effects.  We will not dwell on this issue here, but let us just
mention that they make the kinetic energies $K$ enter as a mass scale
and that they are at the origin of the appearance of log's of $\alpha$
which are characteristic of bound states\footnote{To be more
precise, the logarithms involve the bound state typical energy
scale $\gamma \simeq m \alpha$ and the heavy scale $\Lambda =m$,
leading to a result of the form $\ln(m \alpha / m) = \ln \alpha$.
This log of $m$ is the one we mentioned earlier when talking about
the running of the low energy coefficients. Notice that the
infrared cutoff $\lambda$ always gets canceled while performing
the matching because, by construction, the infrared behavior of the
effective theory is the same as the one of the full theory.}.

Now that we have a way to estimate the order (in powers of $\alpha$)
of the NRQED diagrams contributions, we can go back to the
matching of the effective Lagrangian (Eq.[\ref{nrlag}]) coefficients with QED.
As mentioned earlier, this matching is done by evaluating  scattering
amplitudes evaluated in both theories. There are two expansions 
involved in this matching: an expansion in number of loops in both
QED and NRQED, and an expansion in p/m in NRQED. The number of loops $n$
required is found by simply setting $m \alpha^{3+n}$ equal to the
precision desired in the final result (a factor of $m^3 \alpha^3$ comes
from the wavefunctions and a factor of $1/m^2$ comes from the
nonrelativistic normalization of the Feynman amplitudes). For a given
number of loops, there is an infinite number of nonrelativistic 
interactions in the effective theory but, using the counting rules
described above, only a finite number must be kept for a given
accuracy in the final result.   
Even if the rules were derived for bound state diagrams, they can be
directly applied to the NRQED scattering diagrams {\it i.e.}, one
imagines sandwiching them between wavefunctions and one performs
the same power counting. Once again there are strong ultraviolet
divergences but let us repeat that  the counting rules give us
 the  order at which the diagrams will contribute  once the matching
 of the NRQED coefficients will have been
performed (in other words, once the counterterms will have been
included).

There is only one exception to the counting rules, and it 
occurs when a NRQED scattering diagram  contains a
power-law infrared divergence. In that case, the counting rules
must be modified to take into account the presence of a new scale,
the infrared cutoff. Take, for example, a scattering diagram 
which has    an overall factor of $m^2 \alpha^5$ times an
integral $\int d^3k f(\vec k,\gamma)$. The counting rules would
suggest that this diagram will contribute by a term of  order $m^2 \alpha^5
/ \gamma = m \alpha^4$  to the bound state calculation but,
if it is linearly infrared divergent, it can actually 
 contribute to order $ m^2 \alpha^5 / \lambda = m \alpha^5 \, (m/
\lambda)$. Let us mention that this   type of diagrams 
always contain Coulomb interactions on external legs and the
corresponding integrals are trivial to evaluate. All dependence
on the infrared cutoff are due to the sensitivity of theory to
low momenta. Since, in that region, the effective theory is
``engineered" to reproduce the full theory (here, QED), these
singularities are always present in the scattering
diagrams of both theories so that the renormalized effective
couplings are always $\lambda$ independent.

The use of these rules is better explained with the help of a concrete
example. This is what we do in the next section, 
where we perform an explicit calculation.

\section{Positronium hyperfine splitting}

\subsection{Order $ m \alpha^4$}

In this section, we will apply NRQED to a specific calculation, namely
 the hyperfine splitting\footnote{Which we will denote by
``hfs" in the following.} (E (triplet
state) - E (singlet state)) of ground state positronium. To lowest order,
one has the energy levels of the Bohr theory and there is therefore no
hfs.
 To obtain the first  non-zero contribution to the splitting, one has
to apply first order perturbation theory to the interactions present in the
Lagrangian,  Eq.[\ref{nrlag}]. The
 only interactions that will contribute to the order of 
interest and that are spin-dependent are
\be
 c_1~{ e \over 2m}\,\sigmav\cdot\Bv ,
\ee
which couples to the transverse photon, and the spin-dependent contact
 interaction involving the
electron and positron degrees of freedom (or ``annihilation" diagram)
\be
 d_3 {e^2 \over m^2}\,\chid\sigmav\psi \cdot \psid\sigmav\chi .
\label{cont}
\ee
All the other interactions, because of the presence of additional powers
of $1/m$, will not contribute to this order.

Now we must fix the coefficients $c_1$ and $d_3$ before computing the hfs.
The first coefficient, $c_1$, can be determined by evaluating the
 spin-flipping scattering of an electron from  an external field,
in the extreme nonrelativistic limit. Or one can also simply 
look up the Pauli Hamiltonian,  which the first few terms of  Eq.[\ref{nrlag}],
at tree level,  must reproduce.
 The end result is that $c_1=1$.
 The corresponding Feynman rule, in momentum space, is
\be
\pm e { (\vec p'- \vec p) \times \vec \sigma ) \over 2 m }
\ee
where $\pm$ correspond to the electron and the positron, respectively,
and $\vec p'$, $\vec p$ are the fermion three-momenta after and before
the interaction, respectively. The corresponding contribution to
the hfs is
\ba
&& -i \twowave \nonumber \\ && 
 ~\psi^\dagger [ {e \over 2m} (\vec q - \vec p) \times \vec 
\sigma]_i\psi ~ \chi^\dagger
 [ -{e \over 2m} (-\vec q + \vec p ) \times \vec \sigma]_j
\chi  \nonumber \\ &&  \transv \label{fermisplit}
\ea
 where we have approximated the transverse photon propagator
$i / (p-q)^2 \simeq -i / (\vec p - \vec q)^2 $. Physically, this corresponds
to neglecting the recoil since only momentum, and no energy, is transferred.
The corrections to this approximation are of higher order in $\alpha$.
\par The second term in the square bracket
 is zero because of the triple product.
We can therefore contract the indices $i$ and $j$. We then
  have to spin average the following expression in the ortho and para states:
$$ \psi^\dagger [(\vec p - \vec q ) \times \vec \sigma ]
 \psi \cdot \chi^\dagger[( \vec p - \vec q) \times \vec \sigma ] \chi . $$
 The results are
respectively $ -2/3 (\vec p - \vec q )^2 $ and $ 2 (\vec p - \vec q)^2$.
 Taking the first value minus the second to get the hfs, we
obtain for the integral

\be
 {8 \alpha \pi \over 3 m^2 } \twowave {(\vec p - \vec q)^2 \over (\vec p -
\vec q )^2 }~=~ {m  \alpha^4 \over 3 } .
\ee
Since the $\vec p$ and $\vec q$ integrals decouple, the calculation is trivial,
leading to the famous Fermi splitting\cite{iz}.

Let us now
 consider the annihilation interaction. To fix $d_3$, we once more turn
to the corresponding QED amplitude, namely the tree level annihilation diagram.
In the limit of vanishing external three-momenta, and with nonrelativistic
normalization for the spinors ($\bar u \gamma_0 u =1$ instead
of $2E$), one easily finds that $d_3 =- i e^2/4 $. This interaction
therefore contributes  to the hfs 
\ba
i \psi >< \psi ~&=&~ i \twowave \times 
  \times {-i e^2 \over 4 m^2 }~
\chi^\dagger \vec \sigma \psi \cdot \psi^\dagger \vec \sigma \chi
\nonumber  \\ &&  ~=~
i  {\gamma^3 \over \pi }  \times {-i e^2 \over 4 m^2 }~
\chi^\dagger \vec \sigma \psi \cdot \psi^\dagger \vec \sigma \chi
\label{contact}
\ea
where $\psi >< \psi$ is to be taken as representing the contact
interaction sandwiched between two wavefunctions (see the first
diagram of fig.[4]).
 The spin average of the spinor expression in Eq.[\ref{contact}] gives 2
 in the ortho state
and zero in the para state (as it must  since, by charge conjugation
invariance,  parapositronium cannot decay to an odd number of
photons). The contribution from
the one annihilation kernel is therefore the well known result:

\be
 + { m \alpha^4 \over 4 } .
\ee

\subsection{Order $ m \alpha^5$}
So far, everything has been particularly simple.  The evaluation of the 
$\alpha^5$ corrections,
 however, is a little bit more involved. First, let us notice that,
according to the counting rules given above, it is not possible to identify
any diagram, either in first or second order perturbation theory,
that would lead to a contribution of order $m \alpha^5$. As an example,
consider the annihilation interaction 
 in second order perturbation theory
({\it i.e} two of these interactions sewn together). The wavefunctions
would decouple from the inner loop and lead to a factor of $\vert \Psi 
(0) \vert^2 = (m \alpha)^3/ 8 \pi$. There would be a factor of $(e^2/m^2)^2$
coming from the two contact interactions, and a factor 
of $m$ from the nonrelativistic propagator.  In all we obtain an overall
factor of $ \alpha^5$ times an integral in which the only energy scale
left is
 $ \mu \alpha$. The final result will therefore be of order $m \alpha^6$.
One can convince oneself that no other diagram can lead to a contribution
of order $ m \alpha^5$.

Another 
problem that arises is that, as is well-known from nonrelativistic
quantum mechanics, divergences arise when we apply second order perturbation
theory to the Lagrangian,  Eq.[\ref{nrlag}].

These two difficulties are closely related since they have their origins in 
the fact that we have so far entirely neglected relativistic momenta in our 
analysis. This cannot be correct when one goes beyond first order perturbation
theory since loop momenta are allowed to become arbitrarily high. As explained
above, the correct way to handle this is to renormalize the ``bare"
coefficients of NRQED by matching scattering amplitudes in the effective
theory with QED diagrams.

We will now illustrate this procedure in more details with the contribution
from the annihilation channel, which is represented by 
Eq.[\ref{contact}]
in NRQED.
Renormalizing the coefficient of the contact interaction to
one loop means that we impose the sum of all one-loop NRQED diagrams
involving this interaction to be equal to the corresponding one-loop
QED diagrams, as shown in fig.[4]. As explained above, the counting
rules
tell us that there are no NRQED bound state diagrams contributing to 
order $ m \alpha^6$. We know that the same must hold true for the
NRQED scattering diagrams except if there are infrared
divergent amplitudes. It turns out that there is indeed such
an amplitude here, given by the annihilation interaction followed
by a Coulomb exchange. The corresponding integral is
equal to
$$
\int d^3k~ {1 \over \vec k^2}~ {1 \over \vec k^2 + \lambda^2}
$$
which can easily be seen to be linearly infrared divergent.
It will therefore contribute to order $m^2 \alpha^5  / \lambda$
and must be included in our calculation.

The correction to the lowest
order coefficient of the contact interaction is then equal
 to the one-loop QED scattering amplitudes minus the NRQED diagram
representing the annihilation interaction followed by
 one Coulomb exchange.  It is easy to evaluate the QED
diagrams because the external particles are on-shell and
at threshold. The result is 
\be
  { i  e^4 \over 2 \pi^2 m^2 } - {i e^4 \over 4 \pi} { m \over \lambda}
\ee
for the vertex interaction, after wavefunction renormalization.
Notice the linear infrared divergence. It does not appear in the
result of the one-loop 
renormalized vertex function  $ \Gamma^\mu (p,q)$ in \cite{koniuk} (see
Eq.[A20]). This
is, however,  due to an error in their calculation as can be seen
from the fact that their integral Eq.[A17] does contain a linear
infrared divergence, divergence that cannot be canceled by the
vertex renormalization constant which is only logarithmically
divergent in the infrared.

For the diagram containing the one-loop vacuum polarization,
we find 
\be
{  i e^4 \over 9 \pi^2 m^2} .
\ee
 The
NRQED diagram containing one Coulomb exchange is also easily evaluated
and turns out to be
\be
- { i e^4 \over  4 \pi} { m \over \lambda}
\ee
so that $C^{(1)}$ is
\be
 \biggl[ {i \alpha^2 \over 4 m^2} (32 + {64 \over 9}) \biggr] ~/~ >< 
\ee
where $><$ stands for the (spin averaged) lowest order contact
interaction, namely $-ie^2/ 2 m^2$.

The contribution to the hfs is then, up to one loop, equal to 
\ba
\psi >< \psi (1+C^{(1)})&& ~=~ {m \alpha^4 \over 4} (1 + C^{(1)})
  \nonumber \\ && ~=~ { m \alpha^4 \over 4} - { m \alpha^5 \over \pi} ( 1
+ {2 \over 9})
\ea
which is indeed the correct result\cite{iz}.

\subsection{Order $ m \alpha^6$}
We now briefly outline the calculation of the order $m \alpha^6$
correction due to the one-photon annihilation diagram. The
first thing to do is to renormalize the coefficient of
the NRQED contact interaction to two loops, as we did at the
one-loop level in fig.[4]. Keeping only the terms of order $\alpha^2$,
we obtain the following relation
\ba
>< C^{(2)} &&  + ~C^{(1)} \times {\rm one}~{\rm loop} {\rm ~NRQED~ diagrams}
 \nonumber
\\ && +
{\rm ~ 
two~loops~NRQED~ diagrams}  \nonumber  \\ &&  \equiv 
{\rm QED~two~loops~diagrams}
\label{matching2l}
\ea
where all diagrams are scattering amplitudes
evaluated at threshold and, as mentioned
above, we are only considering corrections to the one-photon
annihilation diagram. One uses once more the counting rules
to determine which NRQED diagrams will contribute to
${\cal O}(m \alpha^6)$ in Eq.[\ref{matching2l}], taking into account
the fact that $C^{(1)}$ is of order $\alpha$.

The rhs (the QED  calculation) in Eq.[\ref{matching2l}] will be of the form
\be
A + B \ln(\lambda/m) + C {m \over \lambda} + D {m^2 \over \lambda^2}
\label{QED}
\ee
where $A,B,C$ and$D$ are constants of order $\alpha^2$. On the other
hand, the NRQED diagrams will lead to a result of the form
\be
A'(\Lambda) + B \ln(\lambda/ \gamma) + C { m \over \lambda} + D {m^2 \over
\lambda^2}
\ee
where $A'(\Lambda)$ contains a finite piece of order $m \alpha^6$ and
ultraviolet divergent terms.
 All
$\lambda$-dependent terms are   reproduced by the NRQED diagrams so
that $C^{(2)}$ is, in the end, infrared finite (but ultraviolet
divergent). The final expression for $C^{(2)}$ is therefore
\ba 
C^{(2)} ~&=&~ \bigl( A -A'(\Lambda) + B \ln(\alpha/2) \bigr) / >< \nonumber
\\ &=&~  - {2 m ^2 \over i e^2} \,\bigl( A -A'(\Lambda) 
+ B \ln(\alpha/2) \bigr) .   \label{uu}
\ea
Notice that, since the NRQED diagrams contain finite pieces, 
 $C^{(2)}$ is {\it not} simply equal to the infrared finite
part of Eq.[\ref{QED}], $A$.

Once $C^{(2)}$ is computed, the correction 
to the hfs is found by calculating
\ba
\psi >< \psi~ (1 +C^{(1)} &  + & C^{(2)} ) \nonumber \\ &  + & {\rm one~} {\rm
 and~ two ~loops ~NRQED
~bound~ state~ diagrams }~\nonumber \\ &=&~{m \alpha^4 \over 4} 
- { m \alpha^5 \over \pi} (1 + { 2 \over 9}) \nonumber \\&&~~~~~
 + B \ln(\alpha/2)
+ {\cal O} ( m \alpha^6)
\label{yaya}
\ea
which is now completely ultraviolet finite. It is easy to understand,
heuristically, why all ultraviolet divergences drop out of
Eq.[\ref{yaya}]. In a nutshell, ultraviolet divergent terms
arise in NRQED for momenta of order $  m$ or greater. Since these
momenta are much bigger than the typical bound state scale
$ \gamma \simeq m \alpha$, the UV divergences in the NRQED
bound state diagrams can be expected to be the same as the
ones encountered in the corresponding scattering diagrams. Since,
in Eq.[\ref{yaya}], each bound state UV divergence is subtracted
from the corresponding NRQED scattering divergence, all $\Lambda$
dependence gets canceled\footnote{Strictly speaking, this is true
only for the power-law UV divergences. The heavy scale {\it is}
(implicitly) present in Eq.[\ref{yaya}], where it cancels the $m$ of
the $\gamma$ factor in the logarithm. See our previous footnote on the
running of the low energy coefficients. For a discussion of the issue
of power-law divergences in a broader context, see \cite{Burgess}.}.

To be more explicit, let us consider in more details one specific
NRQED diagram.
Take, for example,  the bound state diagram containing
the contact interaction iterated twice ({\it i.e.} in
second order of perturbation theory), as depicted in
fig.[5].
It is easy to check that the counting rules predict
a contribution of order $m \alpha^6$ for this diagram. However,
the explicit expression for this diagram is not well-defined
as it diverges linearly:
\ba
\int {d^3p \,  d^3q \,  d^3k \over (2 \pi)^9}& ~& { 64 \pi \gamma^5 \over
((\vec k - \vec  p)^2 + \gamma^2)^2 \, (\vec q^2 + \gamma^2)^2}
({e^2 \over 4 m^2})^2~ \nonumber \\ &&
{ 1 \over - \gamma^2/m - \vec p^2 / m}
\biggl( \chi^\dagger \sigma_i \psi 
 ~ {\rm Tr} ( \sigma_i \sigma_j) ~
 \psi^\dagger
\sigma_j \chi  
\biggr) \biggl\vert_{spin 0}^{spin 1} 
\nonumber \\&=& - { m \alpha^5 \over 4 \pi} { \Lambda / m} + { m
\alpha^6 \over 8} 
\label{ressu}
\ea
where we have put a cut-off $\Lambda$ on the three-momentum
flowing through one of the contact interaction. This divergence
arises from the fact that the Lagrangian Eq.[\ref{nrlag}]
is defined for momenta much smaller than the electron mass
whereas the momentum in a loop is allowed to become relativistic.
 However, for $p \ge m$, one expects the intermediate state loop
 to shrink to a point (in coordinate
space), as far as the low energy theory is concerned.
This means that the divergent contributions to the NRQED
bound state diagrams can be canceled by a renormalization
of the tree level coefficients. In fact, the necessary
counterterms are already present in the coefficients $C^{(1)}$ and
$C^{(2)}$. To see this, let us go back to the definition of $C^{(1)}$,
fig.[4]. When we calculated the $m \alpha^5$ correction, the only
one-loop NRQED diagram we kept in this equation was the
contact interaction followed by a Coulomb exchange. Now, however,
in order to be consistent in our calculation of order
$m \alpha^6$, we must go back to that equation and include
all necessary NRQED one-loop diagrams. One of those diagrams
turns out to be the scattering diagram containing the contact
interaction iterated twice ({\it i.e} fig.[5] with free spinors on the
external legs instead of wavefunctions), 
which has the simple Feynman rule
\be
- \int {d^3k \over (2 \pi)^3} {m \over \vec k^2} 
({e^2 \over 4 m^2})^2~ =~ - 2 { \alpha^2 \over m^2} {\Lambda \over m}. \ee
This leads to a new contribution to $C^{(1)}$ and, consequently, to
a correction to bound state energy equal to (see Eq.[\ref{yaya}])
\be
- \vert \Psi (0) \vert^2 \times - 2 { \alpha^2 \over m^2}
{\Lambda \over m } ~=~ { m \alpha^5 \over 4 \pi} {\Lambda \over m} .
\ee
Combining this with Eq.[\ref{ressu}] yields a finite correction to
the hyperfine splitting equal to  $m \alpha^6/8$.

Before wrapping up this section, we would like to mention that
there is yet another new ingredient appearing at order $m \alpha^6$.
It has to do with the fact that, as we saw earlier,
 the Coulomb interaction does
not increase the order (in powers of $\alpha$) of the bound state
diagrams. 
 Although, in first order perturbation theory this is taken care of 
by the Schr\"odinger wavefunctions, in second order one must also sum
up  the Coulomb interactions in the {\it intermediate} state. In practice,
this is done by using a closed form expression for the full
Coulomb propagator which has been derived in both coordinate and
momentum space \cite{Schwinger}.

\section{Discussion}

We have seen that there are two classes of contributions to bound
state properties, one being characterized by the typical
bound state momentum $ p \approx m \alpha$ and one characterized
by relativistic momenta $p \approx m$. Such a situation is ideal
for the use of an effective field theory (in the decoupling scenario).
NRQED is such a theory, with heavy scale $\Lambda$ equal to the 
electron mass $m$. The use of NRQED greatly simplifies bound state
calculations
 since it involves
only nonrelativistic interactions 
in the bound state diagrams and incorporates the relativistic 
physics via the matching of conventional {\it scattering}
amplitudes. The relative simplicity of the nonrelativistic
interactions makes the bound state diagrams easy to evaluate
whereas the QED interactions appear only in  scattering
amplitudes, which can be evaluated using conventional methods.
This is to be contrasted with traditional
approaches, the Bethe-Salpeter equation for example,
which involve the fully relativistic theory in the
bound state diagrams, leading to the difficulties discussed in Section 2.

We are finally able to understand in what way   the
order $  m \alpha^4$
and $ m \alpha^5$ corrections have a special status:
the $ m \alpha^4$
corrections come exclusively from {\it nonrelativistic} momenta
({\it i.e.} from NRQED diagrams)
whereas the  $ m \alpha^5$ corrections come practically exclusively
from  {\it relativistic} momenta
({\it i.e.} from QED scattering diagrams). We say ``practically" because, as
we saw in the example derived above, there is a linear
infrared divergence in the QED diagrams, showing that
they are in fact sensitive to very low momenta. In fact this
divergence is a pure threshold singularity, which is due to tha
fact that the diagrams are evaluated with external three-momenta
set to zero. However, being a low-energy phenomenon, one expects this
divergence to be reproduced by NRQED and we saw that this is exactly
what happens, in the form of the Coulomb exchange diagram. But aside
from this singularity,  which gets canceled when
matching QED and NRQED, the $ m \alpha^5$ 
contribution originates entirely from QED scattering
diagrams\footnote{Although
we
have explicitly shown this only for the annihilation diagram, this
also holds for the $\vec p \cdot \vec A$ interaction; therefore 
{\it all} ${\cal O} ( m \alpha^5)$ contributions to the hfs are 
purely relativistic.} 
\footnote{It
is in that sense that we call these contributions  ``purely"
relativistic. Maybe a more appropriate terminology would be
``bound state independent". Notice that    what we call ``relativistic
contributions" are sometimes referred to as ``radiative corrections"
in the literature.}.

Let us now compare to the technique of Ref.\cite{koniuk}. 
For the ${\cal O }( m \alpha^4)$ correction, they consider the
QED one-photon exchange in the scattering and annihilation channels,
expanded them to lowest order in $\vec p/m$, and obtained
the expressions Eq.[\ref{fermisplit}] and Eq.[\ref{contact}].
 At this order, this is 
exactly equivalent to using NRQED. In the language of NRQED,
these interactions are found in the effective 
Lagrangian Eq.[\ref{nrlag}], which is obtained by taking the
nonrelativistic limit of the QED Lagrangian. It is in that
sense that we call the  ${\cal O }( m \alpha^4)$ ``purely"
nonrelativistic; they are found directly from the effective
Lagrangian.

Now consider the ${\cal O }( m \alpha^5)$ corrections. Once again,
we restrict ourselves
to the one-photon annihilation  corrections for the sake of simplicity.
In ref.\cite{koniuk}, these corrections are obtained by sandwiching
QED one-loop diagrams between wavefunctions, and taking the limit
$\vec p \rightarrow 0$ in the QED spinors (threshold limit). In that
limit, their calculation reduces to multiplying the one-loop
threshold QED diagrams  by the square of the wavefunctions
evaluated at the origin. This is {\it almost } exactly what we find using
NRQED. Indeed the $( m \alpha^5)$ term comes uniquely from the
renormalization of the NRQED ``bare" parameters, as we demonstrated
explicitly in this paper.
 However, there is an important distinction between the two
calculations. In \cite{koniuk}, the photon mass dependence
was discarded, with no further justification. But in NRQED,
the $\lambda$ dependence gets canceled, order by order in the
matching procedure. Also, it is now easy to understand why
the ${\cal O} ( m \alpha^5)$ result for the ground state can be
trivially generalized to ( $l =0$) states of arbitrary principal
quantum number. Indeed, since the only effect of the bound state
is to provide an overall normalization equal to the square
of the wavefunction at the origin, the ${\cal O} ( m \alpha^5)$
will carry the same dependence on $n$ as $\vert \Psi (0) \vert^2$,
namely it is proportional to $  1/n^3$.

We are now in position to understand how the two approaches
would differ at ${\cal O} ( m \alpha^6)$,
 not only in the handling of the infrared
divergences but also in the prediction of the finite term.
 We saw in Eq.[\ref{yaya}]
that, in contradistinction to the ${\cal O} ( m \alpha^4)$
and ${\cal O} ( m \alpha^5)$ corrections, the $ m \alpha^6$
term contains finite contributions from both relativistic and
nonrelativistic momenta. Notice that this is not  simply a
consequence of the presence of loops since the
${\cal O} ( m \alpha^5)$ term already included integration
over internal momenta. It has for origin a subtle interplay
between relativistic and nonrelativistic flow of momenta through
the Feynman diagrams, interplay which is brought in evidence 
by the use of NRQED and its counting rules. Using the rules of
\cite{koniuk} would yield the relativistic finite 
contribution only (in addition to infrared divergent terms which would
have to be discarded without further justification).

The fact that nonrelativistic momenta  play an important role at this
order also has for consequence that it is no longer 
possible to consider Feynman diagrams containing only a
fixed number of loops, as done in \cite{koniuk}. Indeed, as
already mentioned in Section 5.3, at this order one must consider
bound state diagrams with an infinite number of Coulomb interactions
in the intermediate states.
 For example, were we to complete our calculation of
the ${\cal O} (m \alpha^6)$ contribution due to the 
annihilation interaction in second order of perturbation theory,
we would have to sum up the same diagram with any number of
Coulomb vertices between the two contact interactions. This
complication is directly at the origin of the issue of
reducible {\it vs} irreducible Feynman diagrams in the
Bethe-Salpeter approach. In NRQED, with the help of the Coulomb gauge,
this problem is reduced to its most simple form since it is confined
to the Coulomb interaction and the infinite sum can be performed
using the expressions of \cite{Schwinger}.

\subsection*{Addendum}
While writing this paper we became aware of \cite{Zhang}, in which one
contribution of ${\cal O} (m \alpha^6)$ is calculated.  This 
contribution corresponds to the one-loop vacuum polarization 
correction to the exchange of two photons.

These diagrams, however, lead to purely relativistic corrections
or, in other words, they contribute only to the coefficient $A$ in
Eq.[\ref{uu}]. This is because vacuum polarization is a highly
virtual effect (notice that there is no vacuum polarization
{\it per se} in NRQED; it only enters through the renormalization
of the low energy theory's coefficients). This can be seen 
directly from the fact that the corresponding correction to
the bound state energy is proportional to the square of the
wavefunction at the origin (see Eqs[14-15] in \cite{Zhang}; notice that
$K$ is independent of the bound state momenta $p$ and $q$). In that
respect, the corrections due to these 
 diagrams are similar to the ${\cal O} ( m \alpha^5)$
terms; they are insensitive to bound state physics and can be treated
using the formalism of \cite{koniuk} (especially since they are infrared
finite). An example of a diagram contributing to ${\cal O} ( m \alpha^6)$ 
that would be sensitive to low momenta would be, for example, the three
photon exchange diagrams.
\subsection*{Acknowledgments}
It is a pleasure to thank G. Peter Lepage for teaching me NRQED and
Simon Dub\'e for many useful discussions concerning effective field theories.
This work has been supported by the Natural Sciences and Engineering
Research Council of Canada.


\begin{thebibliography}{99}
\bibitem{koniuk}T. Zhang, L. Xiao and R. Koniuk, Can. J. Phys.
 {\bf 70}, 670 (1992).

\bibitem{NRQED} 
NRQED was introduced in:

W. E. Caswell and G. P. Lepage, Phys. Lett. {\bf 167B}
, 437 (1986).

For  pedagogical introductions, see:

 ``What is renormalization?", G.P. Lepage, invited lectures given at the
TASI-89 Summer School, Boulder, Colorado; ``Quantum Electrodynamics
 for Nonrelativistic Systems and High Precision
Determinations of $\alpha$", G.P. Lepage and T. Kinoshita,
 in {\it Quantum Electrodynamics}, T.Kinoshita
ed. (World Scientific, Singapore, 1990); ``NRQED in bound states: 
applying renormalization to an effective field theory'', P. Labelle,
XIV MRST Proceedings (1992).

\bibitem{BS} E.E. Salpeter and H.A. Bethe, Phys. Rev. {\bf 84}, 1232
(1951); see also review by S.J. Brodsky, in {\it Atomic Physics and
Astrophysics}, edited by M. Chretien and E. Lipworth (Gordon and Breach,
New York, 1969), Vol. I.

\bibitem{NRQCD}  See for example: G.P.Lepage and B.A.Thacker, Nucl.Phys.{\bf B}
(Proc. Suppl.) {\bf 4} (1988), 199;
B.A. Thacker and G.P.Lepage, Phys. Rev. {\bf D 43} (1991), 196;
C.T.H. Davies and B.A.Thacker, Ohio State University preprint DOE-ER-01545-554,
May 1991.;
G.P.Lepage {\it et al}, CLNS preprint 92/1136.

\bibitem{decoupling} T. Appelquist and J. Carazzone, Phys. Rev. D {\bf 11},
 2856
(1975).

\bibitem{iz}  To see the one-photon annihilation  $m\alpha^4$ and $m \alpha^5$
contributions to the hfs in positronium, consult

Itzykson and Zuber, {\it Quantum Field Theory}, McGraw-Hill, 1980, section
10.3.

\bibitem{Burgess} C.P. Burgess and David London, McGill preprint 92/05,
hep-ph preprint 9203216.

\bibitem{Schwinger}E.H. Wichmann and C.H. Woo, J. Math. Phys. {\bf 2},
178 (1961); L. Hostler, J. Math. Phys. {\bf 5}, 591 (1961);
J. Schwinger, J. Math. Phys. {\bf 5}, 1601 (1964).

\bibitem{Zhang} Tao Zhang and Lixin Xiao, Phys. Rev. {\bf A49}, 2411
(1994).

\end{thebibliography}
\end{document}